\documentclass[11pt]{article}
\usepackage{geometry} 
\usepackage{amsmath}
\usepackage{array}
\usepackage{graphicx}
\usepackage[pdftex]{hyperref}
\usepackage[usenames]{color}
\usepackage{fancyhdr}
\usepackage{verbatim}
\usepackage{setspace}
\usepackage{graphicx}
\usepackage{tikz}
\usepackage{pgfmath}
\usepackage{amsfonts}

\newcommand{\myparsum}[1]{}
\title{\textbf{\huge{Mapping the spatiotemporal dynamics of calcium signaling in cellular neural networks using optical flow}}}
\author{\Large{Marius Buibas, $^1$ Diana Yu,$^1$ Krystal Nizar,$^3$ Gabriel A. Silva,$^{1,2,3}$}\\\\
$^1$Department of Bioengineering; $^2$ Department of Ophthalmology;\\ $^3$ Neurosciences Graduate Program, University of California, San Diego\\ 9500 Gilman Drive, La Jolla, CA 92093 USA}

\date{}

\begin{document}
\maketitle{}
{\noindent \textbf{Abstract-}
An optical flow gradient algorithm was applied to spontaneously forming net- works of neurons and glia in culture imaged by fluorescence optical microscopy in order to map functional calcium signaling with single pixel resolution. Optical flow estimates the direction and speed of motion of objects in an image between subsequent frames in a recorded digital sequence of images (i.e. a movie). Computed vector field outputs by the algorithm were able to track the spatiotemporal dynamics of calcium signaling pat- terns. We begin by briefly reviewing the mathematics of the optical flow algorithm, and then describe how to solve for the displacement vectors and how to measure their reliability. We then compare computed flow vectors with manually estimated vectors for the progression of a calcium signal recorded from representative astrocyte cultures. Finally, we applied the algorithm to preparations of primary astrocytes and hippocampal neurons and to the rMC-1 Muller glial cell line in order to illustrate the capability of the algorithm for capturing different types of spatiotemporal calcium activity. We discuss the imaging requirements, parameter selection and threshold selection for reliable measurements, and offer perspectives on uses of the vector data.}\\\

\noindent \textbf{Keywords-} Optical flow; Calcium signaling; Calcium imaging; Neural circuits; Neural networks; Neurons; Glial cells; Astrocytes\\\\

\noindent \rule{6.25in}{0.5pt}
Address correspondance to Dr. Gabriel A. Silva, UC San Diego Jacobs Retina Center, 9415 Campus Point Drive, La Jolla, CA 92037-0946 USA. Electronic mail: gsilva@ucsd.edu

\newpage
\section*{Introduction}

\myparsum{importance of calcium indicators, complexities of calcium signal}
Calcium signaling is an intermediate step in many of the signaling pathways in neurons and glial cells and is informative of functional neural activity. In neurons calcium signaling precedes sub threshold and threshold (i.e. action potential) changes in membrane voltage, and can be used to infer electrophysiology from optical imaging \cite{Smetters:1999p124, Canepari:2006p107, Yaksi:2006p6514, Vogelstein:2009p4525}. In astrocyte glial cells it underlies the mechanisms by which these cells communicate in astrocyte networks and in bi-directional communication with neurons \cite{Agulhon:2007p105, Bennett:2005p102, Scemes:2006p6507}. Relative changes in cytosolic calcium concentration can be measured using different fluorescence indicator dyes that can be imaged by optical microscopy in the visual light range, such as bulk loaded AM esters and genetically encoded calcium indicators \cite{Paredes:2008p7586,Tian:2008p7587}. The emitted fluorescence of indicator dyes change as a function of the relative amount of free calcium ions individual indicator molecules are able to interact with.  Although the relationship between measured fluorescence signals and the calcium levels that produce them is complex and non-linear, it is assumed that there exists a correlation between measured changes in emitted fluorescence by indicator molecules and differing cytoslic calcium concentrations. In this context, the measured fluorescence signal provides a valuable qualitative metric of changing calcium levels that allow inferences of cell signaling and function. Throughout the rest of this paper, we will use the terms ``calcium signal" or ``calcium fluorescence" to mean a measured calcium indicator fluorescence signal that reflects a relative cytosolic calcium concentration, as is routinely implied in the literature, even though in practicality we never know the real, i.e. absolute, free ion concentration that gives rise to the measured fluorescence signal. 

The data collected by a typical experiment records qualitative movies of imaged changes in calcium fluorescence intensity. One can visualize calcium transients and their relative positions and durations, but there is no inherent quantitative analysis of the data by the experiment itself that allows one to derive the dynamics that characterize such signaling events. For example, things such as propagation speeds and directions (i.e. velocity), the kinetics of measured waveforms, or analysis that depend on such properties, such as identifying and mapping the signaling geometry of intercellular calcium waves in networks of neurons or astrocytes. Measuring and tracing calcium (or other second messenger) fluorescence signals quantitatively from recorded movies manually is a tedious and labor intensive process for even small data sets, and involves comparing intensities at different frames and locations in order to calculate speeds and directions. It is generally not possible to do so for large data sets that encompass high spatial and temporal resolution detail or large numbers of cells interacting in a circuit or network.

This can be addressed by analyzing experimental data with a filter algorithm called optical flow, which can be used to derive quantitative measurements of observed spatiotemporal calcium signals imaged from fluorescence movies. The resultant vector data has a variety of uses, ranging from deriving basic measurements of signal velocity and direction, to characterizing and classifying spatiotemporal calcium dynamics between different experimental conditions. Optical flow is an imaging technique (i.e. a filter) that calculates a two-dimensional displacement field between two subsequent frames in a movie, based on the local spatial and temporal gradients of the two images.  The optical flow filter originated in the computer vision field, where it was designed to approximate object motion in time-ordered image sequences for applications like stereo disparity measurements, motion estimation, movie encoding and compression, and object segmentation \cite{HORN:1981p7758}.  The algorithm uses a computed local spatial and temporal gradient to approximate a displacement or flow vector at each pixel in the image. In both neurons and glial cells cytosolic calcium concentration changes manifest themselves as transient responses with a rapid increase, i.e. rising phase, followed by a kinetically slower decaying phase. This is because free calcium is cytotoxic and therefore kept at nanomolar concentrations in the cytoplasm under normal conditions. It is only transiently elevated followed quickly by its re-uptake or extrusion. Temporal changes are typically coupled to spatial changes as a signal propagates through a cell. Measured fluorescence changes then trace specific paths during periods of observation (\textit{c.f.} Fig. \ref{fig:rawframes}). Calcium transients start at a particular location, travel in some direction at a specific speed and terminate at a different location. The typical kinetics of calcium transients in neural cells are particularly well suited to the computational requirements of the optical flow algorithm.

We have successfully applied optical flow to calcium fluorescence movies and obtained displacement vectors that track the spatiotemporal progression of calcium signals.  The filter works for calcium fluorescence data because calcium signals exhibit both spatial and temporal gradients.  The computed vectors provide point estimates of the speed and direction of signals. Optical flow is ultimately an imaging filter that works on whole movies, much like edge filters and image segmentation filters are used in static microscopy \cite{Guo:2004p7781, Mukamel:2009p7794, Hashemi:2008p6111}, and provides a novel and automated way of analyzing the spatiotemporal dynamics of calcium intracellular signaling in neurons and astrocytes.

We begin by briefly reviewing the mathematics of the optical flow algorithm, describe how to solve for the displacement vectors, and how to measure their reliability.  We then compare computed flow vectors with manually estimated vectors for the progression of a calcium signal recorded from representative astrocyte cultures.  Finally, we applied the algorithm to preparations of primary astrocytes and hippocampal neurons and to the rMC-1 Muller glial cell line in order to illustrate the capability of the algorithm for capturing different types of spatiotemporal calcium activity.  We discuss the imaging requirements, parameter selection and threshold selection for reliable measurements, and offer perspectives on uses of the vector data.

\section*{Optical Flow Algorithm and Computation}
\begin{figure}[htbp]
\begin{center}
\includegraphics[width=6in]{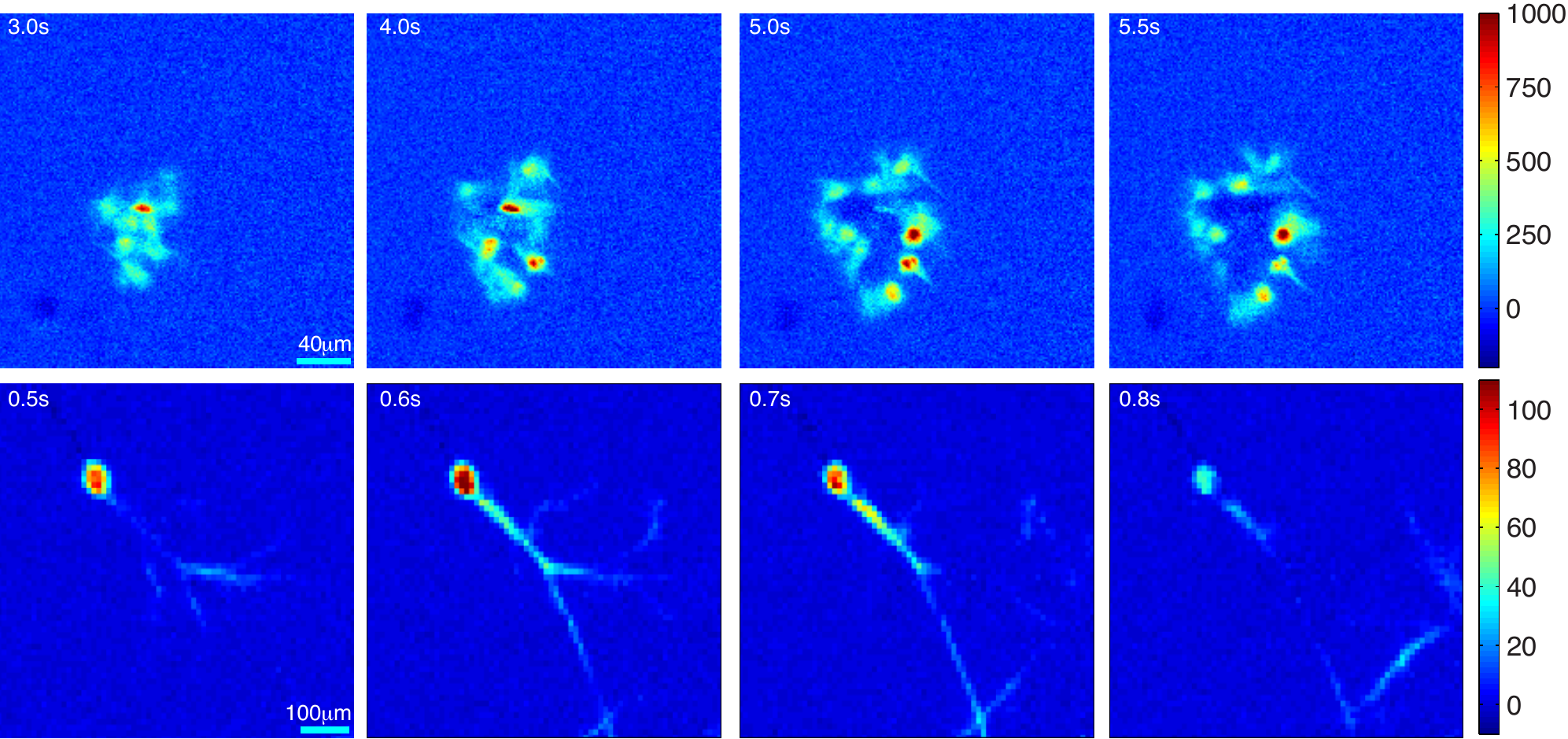}
\caption{Selected frames from recorded movies of imaged calcium fluorescence activity in sparse networks of primary dissociated cortical astrocytes (top) and hippocampal neurons (bottom- where a single neuron at high magnification is shown).  The color coded scale bars on the right represent fluorescence intensity $I$ in units of $\Delta I/sec$, as a first derivative of the calcium signal.  For neurons the signaling was spontaneous, while for the astrocytes waves were pharmacologically induced. Fluorescence increases followed a relatively smooth spatial progression across the frames at the times indicated by the time stamp in the upper left hand corner of each image.  Areas of increasing calcium concentration appear as positive $\Delta I/sec$ values, while areas of decreasing calcium concentration appear as negative values, but at a much smaller magnitudes.}
\label{fig:rawframes}
\end{center}
\end{figure}

In this section we briefly introduce the concepts and mathematics of optical flow, focusing in particular on our own implementation of the algorithm to the experimental data that follows in the Results section. The theory behind the algorithm is well established and the interested reader is referred to a number of excellent texts on the subject (see for example  \cite{HORN:1981p7758, Jahne:2005}). Optical flow is an algorithm that operates at the pixel level and calculates local displacement or velocity between time ordered image pairs. Optical flow (or equivalently image flow) is the perceived motion of an object in a field of view (e.g. by the human eye or a camera), defined as the ``flow" or change in space and time of gray values at the image plane. It is an estimation of the motion field, which is the actual motion of the object in three dimensional space projected onto the image plane (i.e. what we would like to know).  As long as the frequency of successive frames in an image sequence is shorter than the motion or displacement of the object of interest (in order to avoid confounding ambiguities in detecting the components of the motion caused by aperture and correspondence problems- see \cite{HORN:1981p7758, Jahne:2005}), the optical flow algorithm is able to track the motion of objects in the field of view as a function of changing gray scale levels, subject to appropriate constraints and minimizations. In other words, the algorithm assumes that any changes in gray values are due to the object moving, and that the irradiance of the object is constant from frame to frame. (This is actually a weak assumption that is difficult to satisfy since motion usually causes changes in irradiance, which is why the algorithm is an estimation of the motion field. In cases where irradiance does not change, the optical flow exactly equals the motion field.) The algorithm assumes the conservation of gray levels in the field of view and assumes that any changes in the distribution of gray levels are due to motion. In fact, the optical flow constraint equation (introduced below) can be derived by analogy from the continuity equation in fluid dynamics that conserves mass  (\cite{Jahne:2005}). By computing the optical flow for all pixels in a field of view, displacement vectors can be calculated for each pixel that map where an object moved to from the pixel in the first frame to that in the second. Intuitively, one can see why the algorithm performs best with objects that have strong contrasts at boundaries or large signal to background noise ratios. The kinetics of calcium transient signals display clearly distinguishable rising and decaying phases that trace specific paths during periods of observation in the form of intracellular calcium waves (Fig. \ref{fig:rawframes}) that are readily detectable by the algorithm.

\begin{figure}[htbp]
\begin{center}
\includegraphics[width=5.5in]{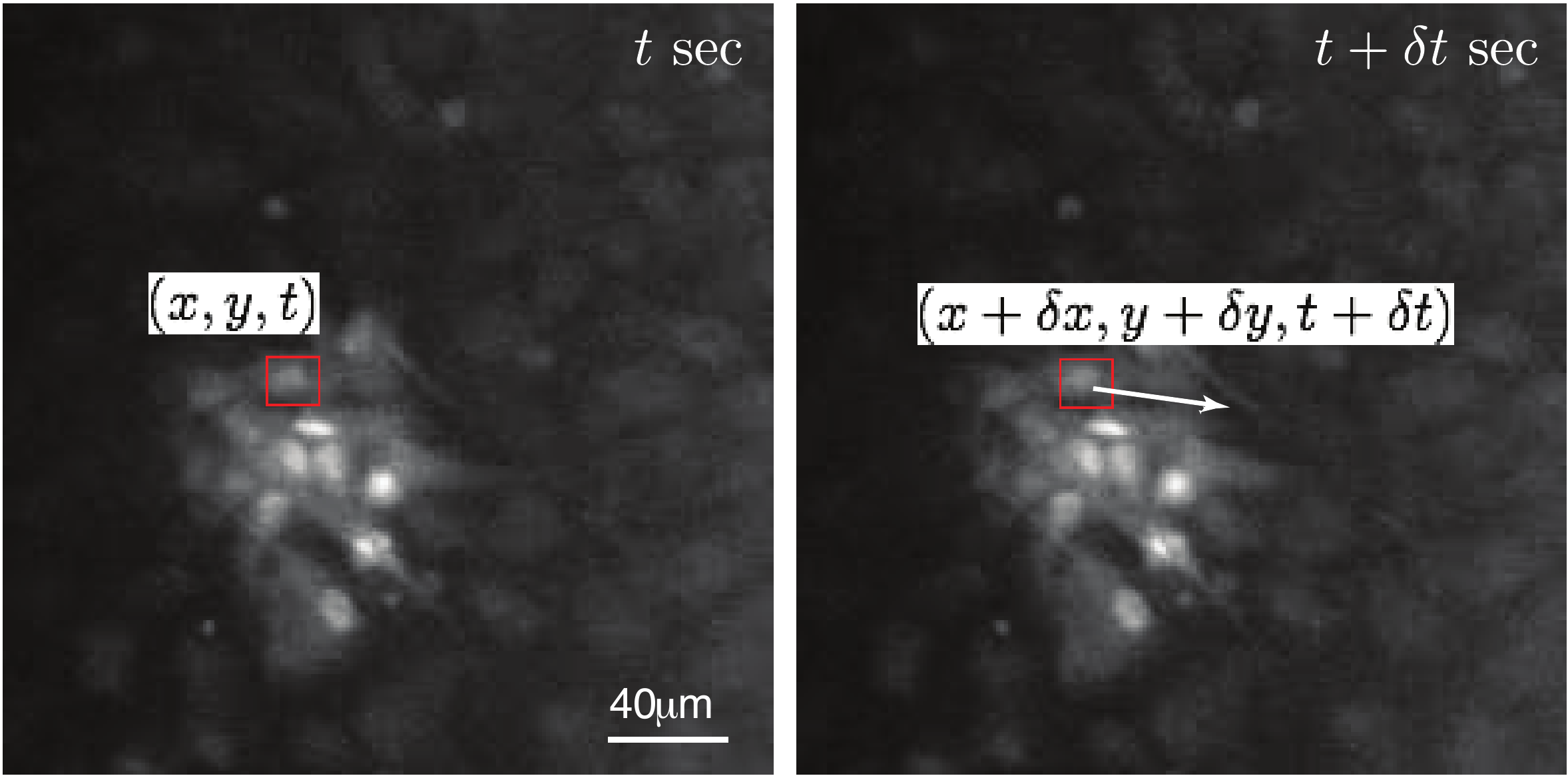}
\caption{The optical flow algorithm.  A window $\Omega$ in the same location in two subsequent image frames is used to compute a displacement or flow vector (arrow) for the pixel at the center of the window.  Only image intensity values in $\Omega$ are used for the calculation.  Vectors are computed for each pixel in an image frame except in border regions where $\Omega$ falls outside of the image. Given the position of the pixel as $(x,y)$ at $t$ seconds,  $(x,y,t)$, the displacement vector defines the motion of the pixel at the subsequent frame at $\delta t$ seconds as $(x+\delta x, y+\delta y, t+\delta t$).}
\label{fig:ofexil}
\end{center}
\end{figure}

The underlying assumption for computation is to constrain local temporal gradients to the product of spatial gradients and displacement vectors. The basic principle of the algorithm takes as inputs two images and computes a vector for each corresponding pixel in the images which approximates the displacement of a small window surrounding that pixel between the two images (Fig. \ref{fig:ofexil}). Only intensity values inside the window are used for computing the pixel displacement value, so the measurement is localized.  Adjacent pixels will have overlapping windows, so their computed vectors will be similar, much like pixels in a blurred image are similar.  Following a mathematical description of the algorithm we describe the method for its solution and implementation that we used to derive the optical flow for calcium signals. We also discuss parameters and constraints of relevance to calcium fluorescence movies.

Consider an arbitrary pixel with gray level intensity $I(x,y,t)$, displaced in the $xy$ plane by $\delta x$ and $\delta y$ at time $\delta t$ in an $n$ x $n$ window $\Omega$ (Fig. \ref{fig:ofexil}). This implies that 
\begin{equation}
I(x,y,t)=I(x+\delta x, y+\delta y, t +\delta t)
\label{eq:Idisc}
\end{equation}

A first order Taylor series approximation of $I(x,y,t)$ by expansion of the right side of \ref{eq:Idisc} results in 
 \begin{equation}
I(x+\delta x, y+\delta y, t +\delta t)= I(x,y,t)+\frac{\partial I}{\partial x}\delta x+ \frac{\partial I}{\partial y}\delta y+\frac{\partial I}{\partial t}\delta t+ \text{higher order terms}
\label{eq:Itaylor}
\end{equation}

Ignoring higher order terms, which provide negligible contributions, and taking into consideration equation \ref{eq:Idisc}
\begin{equation}
\frac{\partial I}{\partial x}\delta x+ \frac{\partial I}{\partial y}\delta y+\frac{\partial I}{\partial t}\delta t=0
\end{equation}

Dividing by $\delta t$
\begin{subequations}
\begin{equation}
\frac{\partial I}{\partial x}\frac{\delta x}{\delta t}+ \frac{\partial I}{\partial y}\frac{\delta y}{\delta t}+\frac{\partial I}{\partial t}=0
\end{equation}
\begin{equation}
\frac{\partial I}{\partial x}u_x+ \frac{\partial I}{\partial y}u_x+\frac{\partial I}{\partial t}=0
\end{equation}
\label{eq:fullconst}
\end{subequations}

The two spatial and one temporal gradients are defined by  $\frac{\partial I}{\partial x}$, $\frac{\partial I}{\partial y}$, and $\frac{\partial I}{\partial t}$, respectively.  $u_x=\frac{\delta x}{\delta t}$ and $v_y=\frac{\delta y}{\delta t}$ represent the $x$ and $y$ spatial components of the optical flow displacement vector $\mathbf{u}(x,y)=(u_x,v_y)$. The basic optical flow formulation is to constrain temporal intensity changes (gradients) to the product of spatial gradients and $\mathbf{u}(x,y)$ to give equation \ref{eq:fullconst}. In more compact notation this can be written as
\begin{align}\label{eq:of}
\nabla I(x,y,t)\cdot \mathbf{u}(x,y)+\frac{\partial I(x,y,t)}{\partial t}=0
\end{align}
Computing optical flow means finding the values of $\mathbf{u}(x,y)$ at each location for every time point that satisfy the above constraint, given the known local image intensity spatial and temporal gradients. 

Two factors establish computability of meaningful non-zero flow vector values.  First, local spatial gradients must be non-zero at the point of interest $(x,y,t)$.  There has to be some image information around the pixel of interest, meaning that neighboring points have to have different values so that gradients are non-zero.  If all pixels in a window around $(x,y,t)$ have the same intensity values, then spatial gradients are zero and motion is undetectable by any means.   Second, for displacement between subsequent frames to be computed, there has to be a temporal gradient at $(x,y,t)$, or some change in intensity between time points.  If there is no temporal change in intensity between subsequent time points, then a value of $\mathbf{u}(x,y)=0$ satisfies the constraint equation in \ref{eq:of}.  Both of these requirements are limitations on the original application of the optical flow when estimating displacement in natural scenes: objects may have constant intensity in a small window and still be moving, meaning that motion may occur and the recorded intensity spatial and temporal gradients equal zero.  These limitations are less important when optical flow is applied to calcium fluorescence movies.

There are many methods for calculating optical flow for recorded movies (see \cite{Barron:1994p6389} for a review), and all of them work on digitized movies with discrete pixel values of position and time, i.e. $(x,y,t) \in (columns, rows, frames)$. We chose the Lucas and Kanade method because it is conceptually simple and efficient, and flexible in terms of the image processing steps required for computation  \cite{Baker:2004p6357,Baraldi:1996p6358,Lim:2005p6397}.  First, computation of the flow vector $\mathbf{u}(x,y)$ is performed on a window or spatial neighborhood $\Omega$ of arbitrary size, centered around $(x,y)$, which is more reliable than a single point estimate at $(x,y)$.  Second, a window function $W(x,y)$ is defined to favor values in the center over those near the edges.  The choice of window size will depend on a variety of factors.  It must be large enough to capture the apparent displacement across frames and small enough to resolve features of interest.  The capture frame rate must be fast enough for displacements to be observable within the width of the spatial observation window across successive frames.  When measuring the spatiotemporal motion of calcium signals the size of the window $\Omega$, the frame rate, and the resolution are all deeply tied to the size of the cells or cellular compartments in which the signal travels.  Together, these parameters must be chosen so that the signal is observable and smooth enough to measure reliably as flow vectors across frames. For example, the choice of parameter values used to image calcium signals in part of a dendrite or a fine astrocyte process will necessarily be different than parameter values for broad calcium signals that fill the soma.The constraint equation is redefined as a weighted least-squares fit of local first-order constraints to a constant model of a local $\mathbf{u}(x,y)$ in each small spatial neighborhood $\Omega$ around the pixel of interest.  The goal is to find the value of $\mathbf{u}(x,y)$ that minimizes  
\begin{align}
\sum_{(x,y) \in \Omega} W^2(x,y)\Big(\nabla I(x,y,t)\cdot \mathbf{u}(x,y)+\frac{\partial I(x,y,t)}{\partial t}\Big)^2
\end{align}
The above equation can be rewritten and solved as the linear system: 
\begin{equation}\label{eq:linsys}
A^T W^2 A \cdot \mathbf{u}(x,y)=A^T W^2 \mathbf{b}
\end{equation}
Where, for neighborhood $\Omega$, consisting of $n$ points centered around the pixel and time of interest $(x,y,t)$, $\Omega=\{(x_{1},y_{2},t), (x_{2},y_{2},t), \ldots , (x_n,y_n,t)\}$: 
 \begin{align}
A&=
\left[
\begin{array}{cccc}
 \frac{\partial I}{\partial x}(x_1,y_1,t) & \frac{\partial I}{\partial x}(x_2,y_2,t) & \dots & \frac{\partial I}{\partial x}(x_n,y_n,t)  \\
 \\
 \frac{\partial I}{\partial y}(x_1,y_1,t) & \frac{\partial I}{\partial y}(x_2,y_2,t) & \dots & \frac{\partial I}{\partial y}(x_n,y_n,t)\end{array}
\right]^{T}
\\
W&=diag\big[W(x_1,y_1),\dots,W(x_n,y_n)\big]\\
\mathbf{b}&=-\Big[\frac{\partial I}{\partial t}(x_1,y_1,t),\dots,\frac{\partial I}{\partial t}(x_n,y_n,t)\Big]^T
\end{align}
$\Omega$ is usually a square window with sizes typically ranging from 3 x 3 to 15 x 15 or $n=9$ to $n=225$ points.  We have set the weight matrix $W$ to a two dimensional Gaussian with $\sigma^2$ equal to 1/6 of the window width.  As an example, for a 5 x 5 or $n=25$ point window:
\begin{equation*}
W=\frac{1}{1000}
\left[
\begin{array}{ccccc}
1 & 6 & 13 & 6 & 1   \\
6 & 54 & 112 & 54 & 6   \\
13 & 112 & 230 & 112 & 13 \\
6 & 54 & 112 & 54 & 6   \\
1 & 6 & 13 & 6 & 1
\end{array}
\right]
\end{equation*}
Here, the center values in $W$ have a greater contribution to the calculation than the edge values, favoring gradient values at the pixel of interest.
 
 Solving for the flow vector $\mathbf{u}(x,y)$ in equation \ref{eq:linsys}, yields: 
\begin{equation}\label{eq:u}
\mathbf{u}(x,y)=\big[A^T W^2 A\big]^{-1} A^T W^2 \mathbf{b}
\end{equation}
Equation \ref{eq:u} describes a linear system in matrix form, where the flow vector $\mathbf{u}$ at spatial and time location $(x,y,t)$ is solved from the quantities of $A$, $W$, and $\mathbf{b}$, defined from the spatial and temporal derivates of $n$ points around $(x,y,t)$.   The $2 \times 2$ matrix $\big[A^T W^2 A\big]$ matrix contains all the image spatial derivatives, and if those values are close to zero, the matrix is poorly conditioned, and flow estimates become unreliable.  Ensuring that both eigenvalues of the $\big[A^T W^2 A\big]$  matrix are sufficiently large is a good way to ensure that the matrix is well conditioned, since a measure of the conditioning number is the ratio of the largest to the smallest eigenvalue \cite{Barron:1994p6389}.  While this is not the only way to  compute conditioning, this is the test we used for visualization and measurement reliability of computed vectors for calcium fluorescence data (see Appendix \ref{ap:eigtest} for more information).

Spatial and temporal derivates were computed using 2 x 2 convolution kernel filters, where the $**$ operator denotes 2-dimensional discrete convolution:
\begin{align}
\frac{\partial I(x,y,t)}{\partial x}&=I(x,y,t)**\frac{1}{4}\left[\begin{array}{cc} -1 & 1\\-1 & 1\end{array}\right]\\
\frac{\partial I(x,y,t)}{\partial y}&=I(x,y,t)**\frac{1}{4}\left[\begin{array}{cc} -1 & -1\\1 & 1\end{array}\right]\\
\frac{\partial I(x,y,t)}{\partial t}&=\frac{1}{\Delta t}\big(I(x,y,t+\Delta t)-I(x,y,t)\big)**\frac{1}{4}\left[\begin{array}{cc} 1 & 1\\1 & 1\end{array}\right] \label{eq:dtcalc}
\end{align}
Here $\Delta t$ represents the time between frames or the frame rate $1/\Delta t$.  Since the temporal derivative calculated in \ref{eq:dtcalc} forms the basis for the $\mathbf{b}$ vector in \ref{eq:u}, the frame rate has a linear effect on the magnitude of the flow vector $\mathbf{u}$.

Optical flow outputs a displacement vector in units of pixels, normalized to the time difference between the two frames used for computation.  When normalized, the vector takes on velocity units of pixels per frame (for this reason it is called a flow vector).  The conversion to physical units will depend on the spatial resolution of the camera and microscope, typically expressed in microns per pixel, and the sampling rate for the movie capture, expressed in frames per second. Spatial resolution is a function of the objectives used as well as the resolution of the imager and any pixel binning used.  The frame rate is limited at the high end by the camera sampling rate, and at the low end by the minimum exposure time required to capture a detectable intensity signal.  The exposure time may be reduced by increased gain or pixel binning, but those come at a cost of reduced resolution or increased noise.  The conversion between units of pixels/frame and units of microns/second is straightforward:
\begin{equation}\label{eq:unitconv}
\frac{microns}{second}=\frac{pixels}{frame} \cdot \frac{frames}{second} \cdot \frac{microns}{pixel}
\end{equation}
While the optical flow algorithm produces vectors in units of pixels/frame, the analysis of the data in the Results section below have been converted into physical units of microns/second, using the resolution and capture rate of the recordings given the specifics of our imaging system.

\section*{Results}
\subsection*{Comparison between computed and manually estimated flow vectors}
\begin{figure}[htbp]
\begin{center}
\includegraphics[width=6in]{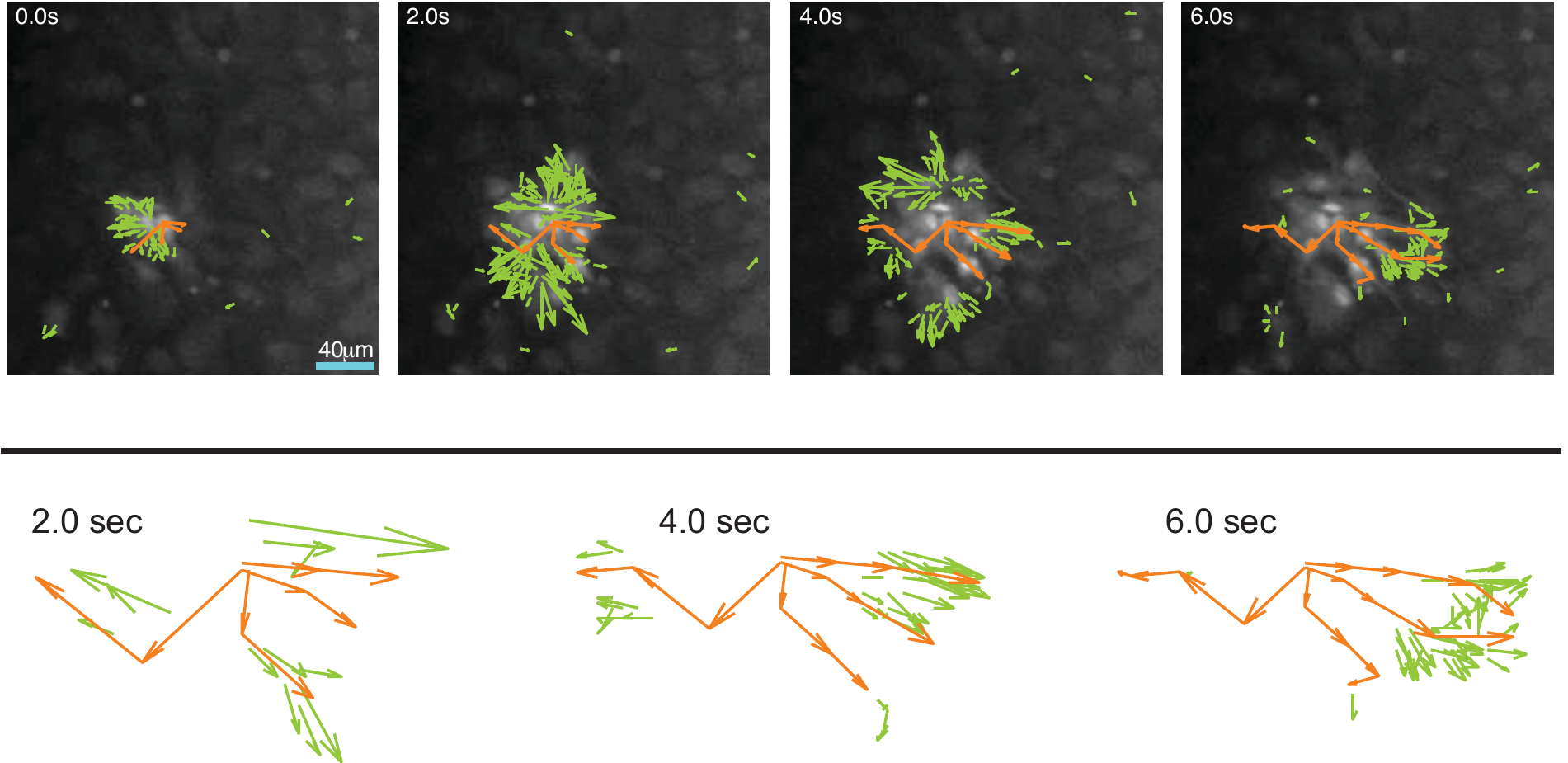}
\caption{Comparison between computed optical flow vectors (green) and manually estimated flow vectors (orange) for a primary dissociated spontaneously forming astrocyte network in culture.  While computed vectors were calculated for every pixel and every frame, manual vectors were estimated every four frames and only trace a few selected signals. Only reliable optical flow vectors are shown, and only one in four vectors in both horizontal and vertical directions are shown for clarity.  Unlike the manual vectors, flow vectors are only shown for the current frame.  The top sequence of panels show vectors overlaid on extracted frames from the actual movie at the indicated times for the entire field of view. The bottom panels show the vectors in detail for the 2, 4, and 6 second frames in order to more clearly assess the qualitative overlap between optical flow computed and manually estimated results. For optical flow vectors (in green) only vectors that putatively correspond to manual vectors (in orange) are shown, in contrast to the upper panels which show all computed vectors (see text). See appendix for details regarding experimental preparations, imaging, and parameters for calculation.}
\label{fig:ofcalc}
\end{center}
\end{figure}

We manually estimated flow vectors for 12 images equivalent to 6 seconds of calcium signaling in primary dissociated spinal cord astrocyte cultures (orange arrows in Fig. \ref{fig:ofcalc}), and qualitatively compared them to computed optical flow vectors for the same data (green arrows in Fig. \ref{fig:ofcalc}; note that only reliable vectors are shown as determined by the eigenvalue test- \textit{c.f.} equation \ref{eq:u} and above discussion). Manual estimation required stepping through frames and approximating roughly how a calcium signal progressed in time, which in this experimental preparation included intercellular calcium waves that propagated through a subset of the cell network.  The manually traced signals were not the only ones observable in the small movie sequence used, but were chosen to illustrate four representative signaling paths.  Manual estimation was performed in two second intervals, estimating the incremental spatial progression of a given calcium signal across four frames. 

Estimation of the flow or displacement of a cell signal such as calcium between frames manually like we did for the data in Fig. \ref{fig:ofcalc} is a very tedious and labor intensive process, and can only realistically be done under very sparse conditions where the observer can clearly delineate the flow of the signal visually. It is nearly impossible to do at the pixel or small window level. In contrast, optical flow calculates a displacement vector for every pixel in every frame, operating at a much finer scale and capturing much more detail than is possible with manual estimates.  Nonetheless, in Fig. \ref{fig:ofcalc} for the purpose of qualitatively validating derived optical flow vectors to manually estimated ones, in both cases there was a clear overlap in vector direction between manual and flow vectors. By contrast, there were greater differences in vector magnitudes between manual and flow vectors, which is consistent with the fact that the manual estimates spanned four frames while optical flow vectors were calculated across adjacent frames. There is also temporal overlap between the two cases, in the sense that similar displacements were estimated for the same frames using both approaches.  The eigenvalue threshold masked out unreliable vectors, and this correlated well with calcium activity; only areas of spatial and temporal changes in the movie produced reliable vectors as assessed visually, which is ultimately the most accurate estimator of complex motions, but only if given the right conditions (e.g. conditions that allow the human eye to separate motion). The optical flow algorithm however, is able to provide reliable quantitative measurements of signaling dynamics at spatial and temporal scales simply not measurable by qualitative visual inspection or manual estimations of the data.

\subsection*{Optical flow characterization of intercellular signaling }
We applied the optical flow algorithm to typical calcium fluorescence movies of spontaneously forming sparse networks of neural glial cells and neurons in culture, and looked at the dynamics of intercellular signaling following pharmacological or mechanical stimulation. We purposely chose sparse networks because it facilitates the visual interpretation of the entire resultant vector field, but the algorithm itself can operate on any data that displays an appropriate signal. We recorded movies from the rMC-1 Muller glial like cell line, which mechanistically displays calcium signaling similar to Muller retinal glial cells \textit{in vivo} \cite{Yu:2009p3681}, primary dissociated spinal cord astrocytes, and primary dissociated hippocampal neurons. Intercellular calcium waves in rMC-1 cells and astrocytes were mechanically induced by gently poking an initial cell without penetrating the cell membrane,  while calcium waves in neuronal networks were induced by the localized pharmacological application of glutamate to one or a small group of cells (see the appendix below for details about experimental preparations and imaging parameters). In particular, intercellular calcium waves in astrocytes and related anatomically specialized macroglial cells such as Muller cells in the neural retina or Bergman glia in the cerebellum have been known to occur under experimental conditions for several years now, and have recently been shown \textit{in vivo} under both physiological and pathophysiological conditions in different parts of the brain, mediated by intracellular calcium transients that induce paracrine signaling, primarily through adenosine triphosphate (ATP) \cite{Kuchibhotla:2009p1752, KurthNelson:2009p4229, Hoogland:2009p4226}. Astrocyte and related macroglial cells engage in bi-directional chemical signaling with neurons and have the ability to modulate and directly participate in information processing in the brain, which necessitates more than just interactions between neurons and almost certainly involves astrocytes somehow. The functional roles of glial intercellular calcium waves and their contributions to modulating neuronal information are not yet known, and in fact the dynamics of these signaling events and the conditions under which they occur are just beginning to be explored.

\begin{figure}[htbp]
\begin{center}
\includegraphics[width=4.8in]{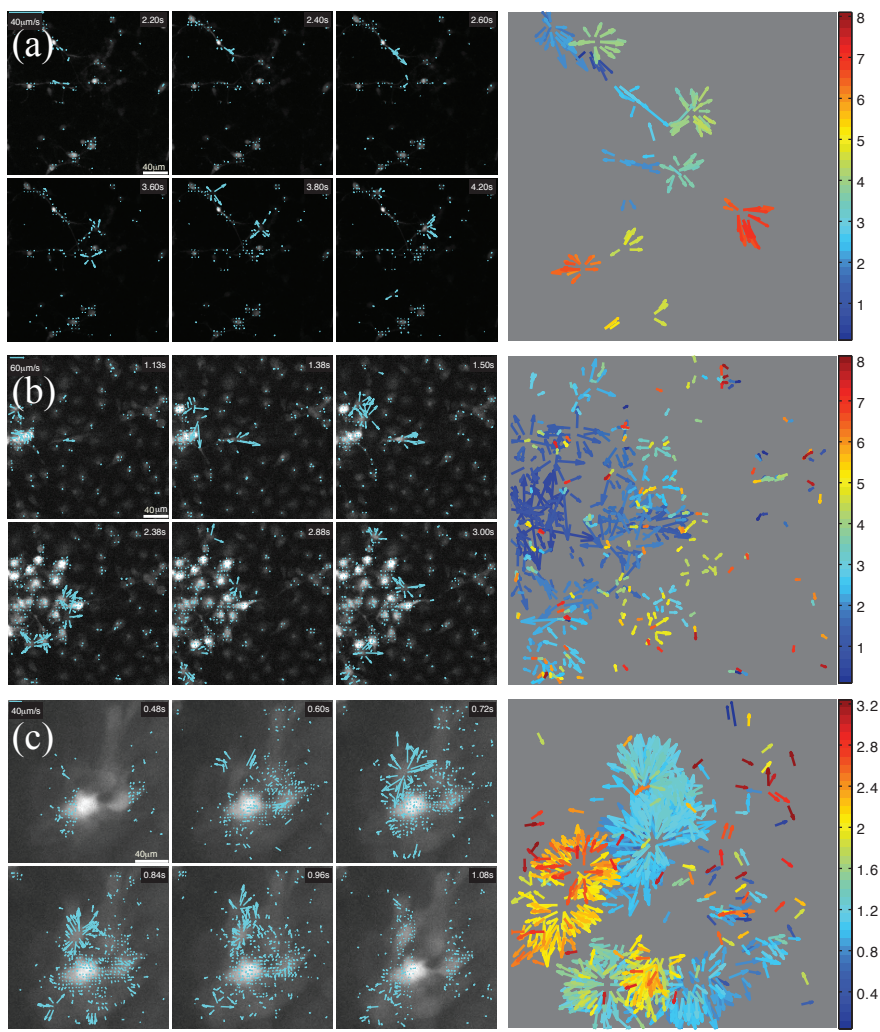}
\caption{Computed optical flow vectors for induced calcium signals in spontaneously forming \textit{in vitro} networks of (a) primary hippocampal neurons, (b) primary spinal cord astrocytes, and (c) the rMC-1 Muller glial-like cell line. Six frames from each representative recorded movie are shown with the computed vector field superimposed at times indicated by the time stamps in each frame (left set of six panels). Right panels: Composite temporal projections of the entire movies. The vector fields show the full spatial progression for the evolving calcium signals, with time (i.e. temporal progression) color coded by the color map (in seconds). Plotting the vector fields in this way allows the full spatiotemporal propagation of derived signals from entire movies to be summarized in a single image. This facilitates the qualitative visualization and identification of complex dynamic signaling patterns that would be difficult to detect otherwise, such as for example by simply "playing back" the movie. }
\label{fig:ofallcells}
\end{center}
\end{figure}

The key parameter for computing optical flow using the Lucas-Kanade method is the window size $\Omega$, specified as a square of a given width (see above).  It defines the local neighborhood of pixels along a point of interest that is used to compute the spatial and temporal gradients required for the calculation. Though not required for computation, a minimum value for the eigenvalues for the matrix $A^{T}W^{2}A$ should be specified to mask out unreliable measurements.  This ensures that only reliable displacement vectors are displayed and used for analysis. Since the intensity values are a function of the experimental setup, microscope, and camera, the $A^{T}W^{2}A$ matrix and its eigenvalues will scale accordingly.  The selection of the eigenvalue threshold is thus arbitrary, much like the selections of the camera gain, exposure time, and other imaging parameters are made to generate easily visible intensity values (see Appendix \ref{ap:eigtest} for more information on selecting suitable eigenvalue thresholds).  Table \ref{tab:params} shows the window sizes, eigenvalue thresholds, and capture frame rates used to calculate the vector fields shown in Fig \ref{fig:ofallcells}.  The displacement vectors can be converted into velocity by equation \ref{eq:unitconv}. The original calcium fluorescence movies and Matlab code written to implement the optical flow algorithm are freely available by contacting the corresponding author.

\begin{table}[htdp]\label{tab:params}
\caption{Image capture and optical flow parameters for shown figures}
\begin{center}
\newcolumntype{C}{>{\centering\arraybackslash}m{.8in}}
\begin{tabular}{|>{\centering\arraybackslash}m{1in}|C|C|C|C|}
\hline
Parameter &  rMC-1 Cells & Astrocytes & Hippocampal Neurons\\
\hline
Frame Capture Rate (Hz) & 16.4 & 8 & 4\\
\hline
Window Size (pixels at 1.3$\mu$m/pixel) & 11 & 9 & 11 \\
\hline
Minimum Eigenvalue - ($\lambda_{1},\lambda_{2}$) greater than & 11 & 1.4 & 0.3 \\
\hline
\end{tabular}
\end{center}
\label{tab:params}
\end{table}

Neuronal cultures displayed derived optical flow vectors along processes as the calcium signal propagated throughout the network. As expected, computed vectors and the resultant vector field followed the geometry of connected processes (i.e. axons and dendrites) in the sparse network (Fig.  \ref{fig:ofallcells}a). The pattern of activation in this example proceeded diagonally from the site of stimulation in the upper left hand corner of the field of view. Some neurons activated at considerably longer times following the stimulus (i.e. out to 7 or 8 seconds) most likely due to recurrent feedback signaling in the network which can last several seconds. Note how since only reliable vectors are plotted, as determined by the eigenvalue test, there are spatial discontinuities in the temporal progression of mapped signals, which reflect areas where the algorithm could not compute reliable vectors given the measured data. This may be especially true at lower magnifications as in the example shown here for comparatively large fields of view that capture many cells. This represents a challenging task for the algorithm. Nonetheless, both the spatial and temporal progression of calcium signals are easily visible. The computed data, being in vector form, can complement existing methods like cross-correlation that use only cell body data to establish relationships between cells for example. 

\begin{figure}[htbp]
\begin{center}
\includegraphics[width=5in]{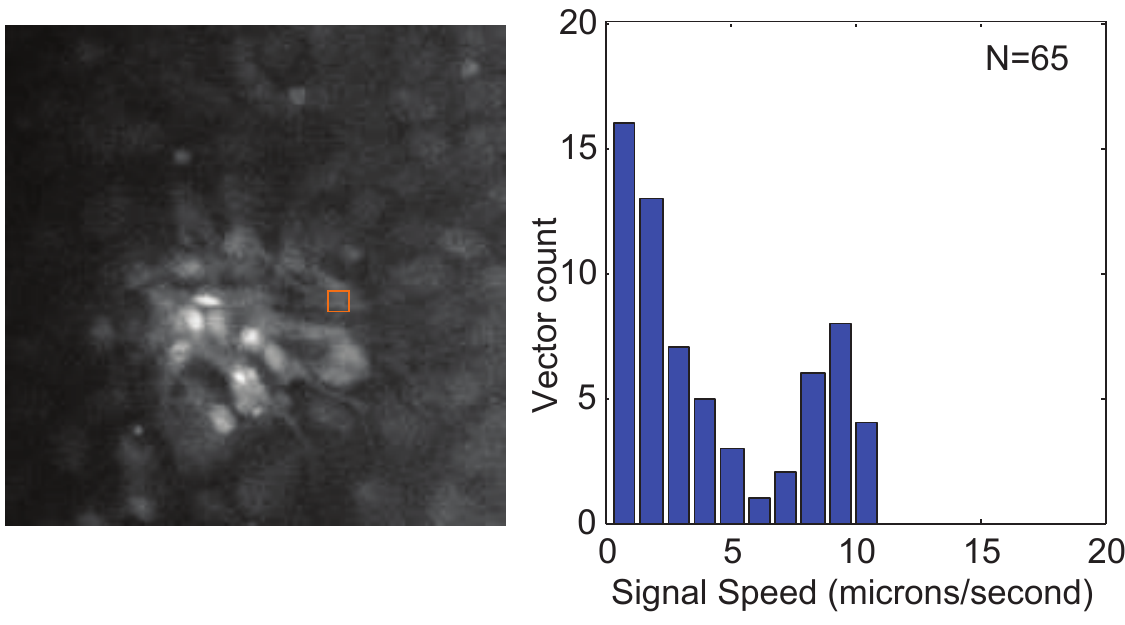}
\caption{Optical flow velocity magnitude distributions for the astrocyte data from Fig. \ref{fig:ofallcells}b. The flow vector magnitudes for reliable measurements in a 10 x 10 pixel (13 x 13 $\mu$m) region (orange square) are shown as histograms.}
\label{fig:ofspeedhist}
\end{center}
\end{figure}

Signal flow patterns were also computed for astrocyte and rMC-1 glial networks (Fig. \ref{fig:ofallcells}b and c, respectively). Astrocyte signaling showed rapid burst like radial patterns that was mostly complete by 2 seconds, with some smaller regions of cells activating later as far out as 6-7 seconds. This is consistent with descriptions of intercellular calcium waves reported previously \cite{CornellBell:1990p5246, Newman:1997p7649, Allen:2005p861, Scemes:2006p1272}. rMC-1 cells showed qualitatively similar radial patterns of activation, with signaling occurring within about 3 seconds following stimulation. However, unlike the astrocyte response, where there was uniform signaling across the network near the site of stimulation, rMC-1 cells showed more heterogeneity in spatial activation patterns, with distinct clusters of cells activating and spreading calcium waves.  The distances traveled by the waves in the rMC-1 example roughly agree with previous quantitative characterizations of calcium waves in similar preparations, on average displaying wave distances of about 60 $\mu$m over the first two seconds or so and distances between 50-100 $\mu$m over about 4 seconds \cite{Yu:2009p3681}. It is interesting however that the spatial progression of the calcium signals in this example was not linear as a function of time, in the sense that cells roughly equidistant from the site of stimulation activated at different times, within about one second for some versus as late as three seconds for others. The relationship and dynamics between the spatial versus temporal properties of such waves are difficult at best and usually not possible to determine by visual inspection of recorded movies alone, and are not captured by calculations such as the one dimensional signaling speed of a progressing wave front. Furthermore, speed and distance calculations of neuronal and glial signaling across networks of cells are usually corse approximations computed using low magnification movies that provide a sufficiently large field of view. In contrast, optical flow provides reliable single pixel vectors for any sized region of interest that represent very fine grain detailed descriptions of calcium signal propagation difficult to achieve otherwise. For the astrocyte data from Fig. \ref{fig:ofallcells}b, Fig. \ref{fig:ofspeedhist} illustrates the distribution of signaling speeds in $\mu$m/second for 65 optical flow vectors for a small 10 x 10 pixel region equivalent to a 13 x 13 $\mu$m region in the field of view (orange box in the figure). Any region size of interest anywhere in the imaged field that might be of functional interest to the investigator can be similarly characterized. By way of rough comparison, optical flow calculated speeds for calcium signals in computed window were distributed from 1-10 $\mu$m/sec, and are roughly similar to those reported previously using more approximate methods, in the range of 5 to 10 $\mu$m/second \cite{CornellBell:1990p5246, Newman:1997p7649, Allen:2005p861, Scemes:2006p1272}. The bimodal distribution in the figure reflects what is visually apparent in the source movie:  some of the areas in the orange pixel region exhibit spatiotemporal displacement while others do not, indicating that calcium concentration changes propagate in specific regions with specific patterns.  Manual estimates from the literature typically look at maximum propagation speeds, as seen in the second peak at about 9 $\mu$m/second.

\section*{Discussion}
We describe and show the application of optical flow gradient methods for identifying and spatiotemporally mapping functional calcium signaling in networks of neurons and glia.  Although we focused on networks of cells here, the method can be equally applied to the analysis of spatially detailed sub-cellular compartmentalized regions of interest, such as dendrites or astrocyte processes. The method makes use of the spatial first derivative of moving objects in a field of view, in this case changes in fluorescence levels of calcium indicator dyes associated with the free concentration of intracellular calcium, to track their motion between subsequent frames in an image sequence (i.e. a recorded movie).  The mathematical foundations of optical flow are well established and optical flow algorithms have been used in a wide variety of fields including applications to cell and molecular biology to track the movement of proteins, vesicles, and even whole cells \cite{Miura:2005p6429}.  In neuroscience and neural engineering it has been used in electromyography \cite{Knuttinen:2002p6395,Ostlund:2007p6430} and sensory perception \cite{Langley:2007p6396,Pagano:1998p6436}, while clinically it has been used to detect seizures in neonatal infants \cite{Karayiannis:2006p6394}, among other applications.  However, the method has not been previously applied to tracking and visualizing calcium signaling and deriving quantitative measurements of calcium spatiotemporal changes that underlie intracellular and intercellular functional signaling in neural cells.

Although in this paper we applied the optical flow algorithm to two dimensional fluorescence movies, the algorithm itself can be readily applied to a recorded movie made up of three dimensional stacks acquired using two-photon microscopy. Work by others is pushing two photon imaging towards recording real time functional signaling from three dimensional volumes of active cellular neural networks \cite{Gobel:2007p137, Gobel:2007p1640}.  If the sampling rate is sufficiently high, optical flow can be computed in three dimensions using a volume instead of a square window around a pixel to generate a three dimensional displacement vector.  The same constraints on volume size, sampling, and vector reliability metrics  in two dimensions apply to the three dimensional case.

Optical flow methods produce a lot of data, generating a vector for every pixel in every image pair computed, so further processing, rendering and visualization methods are key to making quantitative comparisons between experimental setups.  Statistical comparisons can be made from vector values by comparing differences between selected regions in different preparations; velocity averages for each region can be compared using statistical methods such as means, standard deviations, and p-values.  While vector values from adjacent pixels are not statistically independent, averaged vector values for a given region of interest may be used for statistical comparison with another, non-overlapping region.

Another potential use of the vectors is to classify spatiotemporal patterns.  Similar to using a scalar kernel filter to match an image pattern such as an edge or corner, vector fields themselves can be filtered with a known vector kernel to match a pattern of interest.  This method is called Clifford convolution \cite{Ebling:2003p8085} and has been used to label physical flow regimes in fluid dynamics applications.  By designing a vector field filter and convolving it with computed optical flow vectors, a scalar map identifying specific patterns of flow associated with the saptiotemporal dynamics of the measured signal can be constructed in order to classify regions exhibiting such patterns.

One of the most exciting potential uses of computed flow vectors is in functional network reconstruction.  Borrowing again from the field of fluid mechanics, a dynamic vector field can be used to reconstruct the path of a hypothetical particle from a given starting point, tracing out the path that a signal might take between cells, much like a particle in a dynamic flow field \cite{Weiskopf:2005p8075, Weiskopf:2005p8065}. Geometrically mapped paths of measured signals that originate in an activating cell and propagate through a network may be very useful for reconstructing the dynamics of the network. This would complement existing network reconstruction algorithms which typically rely on temporal data around fixed regions of interest. 

\subsection*{Acknowledgments}
This work was supported by grant RO1 NS054736 from the National Institute for Neu-rological Disorders and Stroke (NINDS) at the National Institutes of Health (NIH).

\bibliographystyle{plain}

 \bibliography{./optiflow}

\appendix{}
\section{Appendix}
\subsection{Cell Preparations}
rMC-1 glial cells and primary spinal cord astrocyte cultures (the latter dissected and grown similar to previously described \cite{Silva:1998p523, MacDonald:2008p1524} were grown to approximately 80\% confluency and washed twice with Kreb-HEPES buffer (KHB) solution (10 mM HEPES, 4.2 mM NaHCO$_{3}$, 10 mM glucose, 1.18 mM MgSO$_{4}$, 7H2O, 1.18 mM KH$_{2}$PO$_{4}$, 4.69 mM KCL, 118 mM NaCl, 1.29 mM CaCl$_{2}$, pH 7.4) and incubated with 5$\mu$M Fluo-4AM in KHB for 1 hr at room temperature. Excess dye was removed by washing twice with KHB and an additional incubation of 30 min at room temperature was done to equilibrate intracellular dye concentration and ensure complete intracellular hydrolysis. For astrocytes and rMC-1 cells, calcium transients were initiated by mechanical stimulation of a single cell using a (0.5$\mu$m i.d.) micropipette tip (WPI Inc., Sarasota FL) mounted on a M325 Micrometer Slide Micromanipulator (WPI Inc., Sarasota FL). Comparable data were obtained using adenosine triphosphate (ATP) pharmacological stimulation.

For hippocampal cultures, dissociated hippocampal neurons from timed-pregnant embryonic day 18 (E18) Sprague-Dawley rats were cultured on glass bottomed tissue culture dishes coated with poly-D-lysine and laminin (BD Biosciences, San Jose, CA). Cultures were plated at a cell density of $10^{6}$ cells/3.8cm$^{2}$. Cultures were maintained at 37C in 5\% ambient CO$_{2}$. Plating media was composed of basal medial Eagle (Invitrogen, Carlsbad, CA) with 1X Glutamax, 1000 U/mL penicillin and streptomycin sulfate, 5\% FBS, and 1X N2 supplement. Culture media consisted of Neurobasal (Invitrogen, Carlsbad, CA) with 1X Glutamax, 1000 U/mL penicillin and streptomycin sulfate, 20mM glucose, and 1X B27 supplement. Culture media was supplemented with 10uM Ara-C for 24 hrs at 1DIV to inhibit overgrowth of glia. All imaging was performed on 3-5DIV.

Bulk loading of hippocampal cell cultures was accomplished via incubation in the dark, at room temperature, for 30 min in 1$\mu$M of the fluorescent Calcium indicator Fluo-4-AM in Krebs-HEPES buffer (10mM HEPES, 4.2 mM NaHCO$_{3}$, 10mM glucose, 1.18mM MgSO$_{4}\cdot$7H$_{2}$O, 1.18mM KH$_{2}$PO$_{4}$, 4.69 mM KCl, 118mM NaCl, 1.29 mM CaCl$_{2}$, pH 7.4), followed by 2x 5 min washes in Krebs-HEPES with 100$\mu$M sulfinpyrazone. Hydrolysis was allowed to proceed for an additional 30 min. Stimulation of neurons with glucose was performed by microinjection of 100uL of 10mM glutamate in PBS from a specified-side of the culture dish, well outside of the microscope field of view. The fluorescence signal generated across the monolayer of cells was recorded for 10 sec prior to glutamate injection, and for 120 sec following injection. Cultured neurons were incubated for 30 min prior to imaging in Mg$^{2+}$-free PBS to induce the synchronization of calcium transients.

\subsection{Imaging Setup}
Visualization of calcium indicator dye fluorescence was achieved using a 488 nm (FITC) filter on an Olympus IX81 inverted fluorescence confocal microscope (Olympus Optical, Tokyo, Japan) that included epifluoresence, confocal, phase, brightfield, and Hoffman differential interference contrast (DIC) modalities. Real-time movie recordings of calcium transient propagation were acquired with a Hamamatsu ORCA-ER digital camera (Hamamatsu Photonics K.K., Hamamatsu City, Japan) and Image-Pro Plus data acquisition and morphometric software (version 5.1.0.20, Media Cybernetics, Inc., Silver Spring, MD) or LabView custom written data acquisition software (ScopeController). All images were captured with a 10X objective, using a 2x2 binning on the camera, for a resolution on 1.3$\mu$m/pixel and a total image size of 612x572 (camera's maximum resolution is 1224x1144). Images sampled at frequencies ranging from 2 to 16.4Hz, or 0.5sec to 0.06sec exposure time.

\subsection{Reliable Vectors via the Eigenvalue Test}\label{ap:eigtest}
Recall that flow vectors are computed from the linear system in \ref{eq:u}.  This is a typical linear system of the form:
\[ M\cdot \mathbf{u}=\mathbf{z} \]
where $\mathbf{u}$ is the unknown and $\mathbf{z}$ and $M$ are known quantities.  The condition number of a matrix simply describes how a small deviation in the known $\mathbf{z}$ translates to an error in $\mathbf{u}$.  A high condition number means the matrix is ill-conditioned, meaning that a small deviation in $\mathbf{z}$ leads to a large deviation in $\mathbf{u}$, making that computation unreliable.  One way to compute the condition number of a matrix is to take the ratio of the largest to smallest eigenvalue of that matrix:
\[ \kappa(M)=\left | \frac{\lambda_{max}(M)}{\lambda_{min}(M)} \right | \]
Since $M$ is a $2\times2$ matrix, it has 2 eigenvalues so ensuring that both are above a certain value makes the condition $\kappa$ value relatively low.  The minimum threshold value depends on the incoming intensity values.

Intensity readings from the CCD camera can take on any number of values, based on the digitization (8-bit, 12-bit, 16-bit, for example), the exposure time, gain setting on the camera, and above all the dye loading in the cell preparation.  Typically, during observation, the experimenter manually adjusts gain and exposure time to obtain reasonable intensity values, typically in the middle of the digitization range.  

Eigenvalues for the $A^{T}WA$ used in flow vector calculation typically scale with the range of recorded intensity values and are calculated for every pixel, producing an eigenvalue image map.  The values chosen in table \ref{tab:params} were manually chosen during examination of the minimum eigenvalue image for a few representative frames, ensuring that they fell between ares where we visually detected spatiotemporal changes in intensity and areas were we did not detect such changes.  This is the same process one would undertake when thresholding a regular monochrome image for the counting of cells: the intensity threshold is set to a value between the intensity of an area where there is a cell and an area where there is no cell.

\end{document}